\documentclass{aip-cp}
\usepackage[numbers]{natbib}
\usepackage{graphicx}
\usepackage{amssymb}
\usepackage[english]{babel}

\begin{document}

\title{Dynamics of Coherent States in   Regular  and Chaotic Regimes of the Non-integrable Dicke Model}

\author[aff1,aff2]{S. Lerma-Hern\'andez\corref{cor1}}
\author[aff2]{J. Ch\'avez-Carlos}
\author[aff3]{M. A. Bastarrachea-Magnani}
\author[aff2]{B.  L\'opez-del-Carpio}
\author[aff2]{J. G. Hirsch} 


\affil[aff1]{Facultad  de F\'\i sica, Universidad Veracruzana, Circuito Aguirre Beltr\'an s/n, C.P. 91000  Xalapa, Mexico}
\affil[aff2]{Instituto de Ciencias Nucleares, Universidad Nacional Aut\'onoma de M\'exico, Apdo. Postal 70-543, C.P. 04510  Cd. Mx., Mexico}
\affil[aff3]{Physikalisches Institut, Albert-Ludwigs-Universitat Freiburg, Hermann-Herder-Str. 3, Freiburg, Germany, D-79104.}
\corresp[cor1]{Corresponding author: slerma@uv.mx}

\maketitle
\begin{abstract}
The quantum dynamics of initial coherent states is studied in the Dicke model and correlated with the dynamics, regular or chaotic, of their classical limit. Analytical expressions for the survival probability, i.e. the probability of finding the system  in its initial state at time $t$, are provided in the regular regions of the model. The results for  regular regimes are compared with those of the chaotic ones. It is found that initial coherent states in regular regions have a much longer  equilibration time than those located in chaotic regions. The properties of the distributions for the initial coherent states in the Hamiltonian eigenbasis are also studied. It is found that for regular states the  components with no negligible contribution are organized in  sequences of energy levels  distributed according to Gaussian functions. In the case of chaotic coherent states, the energy components do not have a simple structure and the number of participating energy levels is larger than in the regular cases.     
\end{abstract}

\section{Introduction}

Simple models, originally proposed to study schematically  quantum phenomena present in more realistic and elaborated systems, have become experimentally feasible. Examples of these include the kicked rotor \cite{KR-T}, the Lipkin-Meskov-Glick  \cite{LMG} and the Dicke model \cite{Dicke54,Hepp73}, whose experimental realizations have been reported, respectively,  in Refs. \cite{KR-E}, \cite{Ober10} and \cite{Bau10}. In this contribution we focus on the latter one, the Dicke model, and present recent results we have obtained   by studying the onset of chaos  in the model.  

The Dicke model was proposed to describe the interaction between a set of $N$ two-level atoms with a single mode of the electromagnetic field. Its Hamiltonian reads (we set $\hbar=1$ throughout all the paper)
\begin{equation}
\hat{H}_{D}=\omega \hat{a}^{\dagger}\hat{a}+\omega_{0}\hat{J}_{z}+ \frac{\gamma}{\sqrt{N}}\left(\hat{J}_{+}+\hat{J}_{-}\right)\left(\hat{a}+\hat{a}^{\dagger}\right),
\label{Dicke}
\end{equation}
where $\hat{J}_i$   are atomic  pseudo-spin operators satisfying the  $\textrm{su}(2)$ algebra,   $\hat{a}$ $(\hat{a}^\dagger)$ is the field bosonic annihilation (creation) operator,   and $\gamma$ is the coupling strength between the field and the  atoms.
  The pseudo-spin operators are defined as
 \begin{equation}
 J_z= \frac{1}{2}\sum_{i}^{N} \left(|e_i\rangle   \langle e_i|-|g_i\rangle \langle g_i| \right), \ \ \ \ 
 J_+= \sum_{i}^{N} |e_i\rangle   \langle g_i|, \ \ \ \ 
 J_-= \sum_{i}^{N} |g_i\rangle   \langle e_i|, 
 \end{equation}
  where $|g_i\rangle$ and $|e_i\rangle$ are   the ground- and excited-state of the $i$-th atom, respectively. 
 The $\textrm{su}(2)$ algebraic structure of the model simplifies it strongly. The whole Hilbert space is divided in  invariant subspaces labelled by the pseudo-spin quantum number $J$,  with the ground-state belonging to the subspace with the  largest value ($J=N/2$). 

The Dicke model exhibits a quantum phase transition (QPT) \cite{DickeQPT} at a critical value of its  coupling constant  $\gamma_{cr}=\sqrt{\omega\omega_o}/2$. The critical value separates  a normal phase ($\gamma\leq \gamma_{cr}$) from a superradiant  phase ($\gamma>\gamma_{cr}$).  Other critical phenomena, the so-called excited-state quantum phase transitions (ESQPT), have also been studied in  the Dicke   model \cite{Bas14},  the main signature of their presence is a singularity in the density of states  \cite{ESQPT}.      

The  Dicke Hamiltonian has an additional  discrete parity symmetry, which separates the $J$ sectors of the Hilbert
space in two invariant subspaces,  they are spanned  by
$$
\mathcal{H}_+=\{|J,m\rangle\otimes|n\rangle | m+J+n {\hbox{  even}} \}
\ \ \ \ {\hbox{ and }} \ \ \ \ 
\mathcal{H}_-=\{|J,m\rangle\otimes|n\rangle | m+J+n {\hbox{  odd}} \},
$$   
respectively. Due to the parity symmetry, in the superradiant phase there are  pairs of degenerate energy states,   for all the energy eigenvalues in the  interval ranging from the ground-state to the critical energy of the ESQPT~\cite{ARMANDO}. To avoid complications related to these degeneracies, we consider here only the parity positive sector. 

\section{Classical chaos and its quantum signatures }
Chaos is a classical concept that implies  extreme sensitivity to initial conditions. For chaotic Hamiltonian systems, two trajectories initially separated a distance $d_o$ in the phase space will separate exponentially in the tangent space as a function of time, with a rate given by 
$$
d\approx d_o \exp(\lambda t), 
$$
where $\lambda$ is the maximal Lyapunov exponent \cite{Lyapunov}. The previous concept does not have  a direct translation to the quantum realm, where  trajectories in  phase space are not defined. However, since the classical description must emerge from the quantum theory, a natural question arises:  what are the signatures of  chaos in a quantum system whose classical limit is chaotic?  Some partial answers to the previous question includes the Bohigas-Giannoni-Schmit conjecture \cite{BGS}, which establishes that the nearest neighbour spacing distribution of the energy levels in a quantum system with a  chaotic classical limit is described by random matrix ensembles, the Gaussian orthogonal ensemble in the particular case of time reversal invariant   Hamiltonians, as is the case of the Dicke Hamiltonian.
  
\subsection{Classical dynamics}
A way to define the classical limit of the Dicke Hamiltonian relies on the coherent states $|z, \alpha\rangle=|z\rangle\otimes | \alpha\rangle$, where  
($z, \alpha \in \mathbb{C}$)
$$
|z\rangle=\frac{1}{\left(1+\left|z\right|^{2}\right)^{j}} e^{z \hat{J}_+}|J, -J\rangle \  \ \ \ \ \ {\hbox{and}}\  \ \ \ \ \ 
|\alpha\rangle=e^{-|\alpha|^2/2}e^{\alpha \hat{a}^\dagger}|0\rangle,
$$
for the pseudospin  and bosonic sector, respectively. 
From this set of states (forming an overcomplete basis for the Hilbert space of states), a classical Hamiltonian can be defined through the expectation value
\begin{equation}
H_{D}^{cl} = \langle z, \alpha | H_D|z, \alpha\rangle 
 =  \frac{\omega}{2}\left(p^{2}+q^{2}\right)+\omega_{0}\,j_{z}+2\gamma\sqrt{J}\sqrt{1-\left(\frac{j_{z}}{J}\right)^{2}}\,q\,\cos\phi,
\end{equation}
where the canonical variables $(\phi,j_z)$ and $(q,p)$ are related to  the coherent state parameters by
$$
z=\sqrt{\frac{1+j_z/J}{1-j_z/J}}e^{-i \phi}\ \ \ \ {\hbox{  and }} \ \ \ \
\alpha=\frac{q+ip}{\sqrt{2}},
$$
respectively.  The classical phase space of  the previous Hamiltonian is $\mathbb{R}^2\times \mathbb{S}^2$, where the real plane comes from the bosonic part and the two-dimensional surface of a 3D-sphere of radius $J$ comes from the pseudospin part.   It can be shown  \cite{coreanos} that the previous classical Hamiltonian gives the leading order contribution to the quantum propagator in the  limit  $J\rightarrow \infty$, the effective Planck constant being $\hbar_{eff}=1/J$.

 Coherent states satisfy   minimal uncertainty relations in the   phase space \cite{CSbos}. Recalling we are assuming $\hbar=1$,  for the bosonic part they satisfy $\Delta \hat{q}\Delta \hat{p}=1/2$ (for a general state $\Delta \hat{q}\Delta \hat{p}\geq 1/2$), where $\hat{q}$ and $\hat{p}$ are the quadratures of the bosonic operators [$\hat{q}=(\hat{a}+\hat{a}^\dagger)/\sqrt{2}$ and $\hat{p}=i(\hat{a}^\dagger-\hat{a})/\sqrt{2}$], whereas for the pseudospin  case  the coherent states minimize the sum of uncertainties \cite{CSpseu}  $\Delta^2\equiv \Delta J_z^2 +\Delta J_x^2 + \Delta J_y^2=J$  [in general for any state  $J\leq \Delta^2 \leq J(J+1)$].  Because of the previous properties,   coherent states are considered the  quantum states   closest to  classical states \cite{CSBook}, the latter ones being represented by points in phase space  with zero uncertainties.    

In the following, we  discuss some recent results that we have obtained by investigating the presence of chaos in the previous classical model and the corresponding signatures we have identified in its quantum version  for finite (but relatively large) $J$. To enhance the correspondence between quantum and classical results,  we employ coherent states as trial states in our quantum studies.  

\subsection{From regular to chaotic dynamics: classical and quantum signatures}

\begin{figure*}
\begin{tabular}{lcccr}
 & & \includegraphics[width=0.22\textwidth]{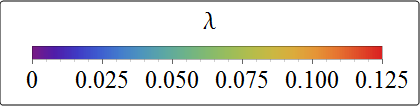}&\includegraphics[width=0.22\textwidth]{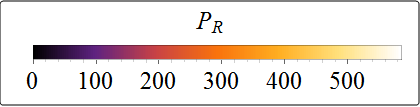}& $E/J$ \\
 & \includegraphics[width=0.26\textwidth]{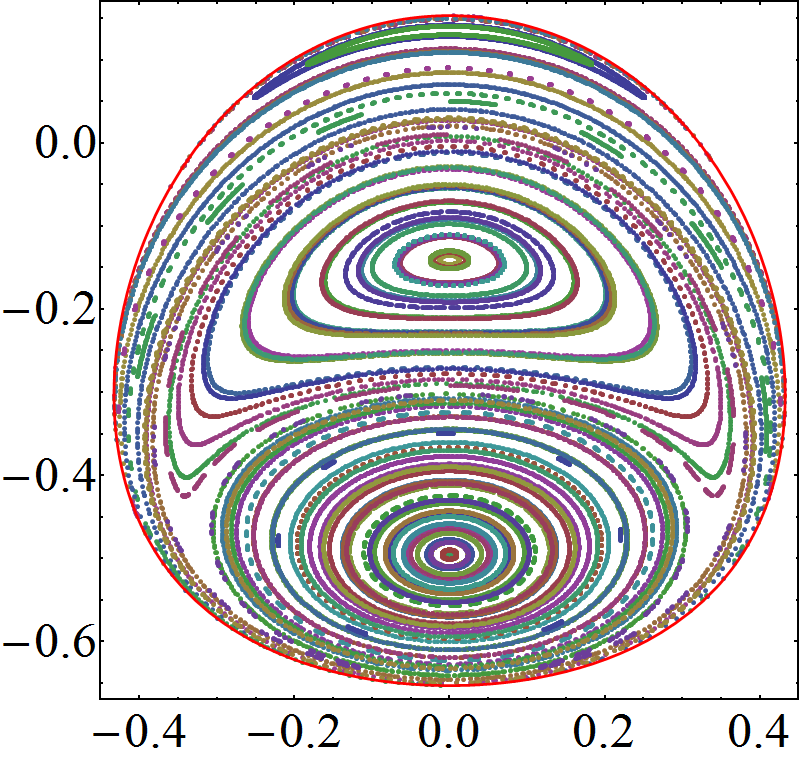}&\includegraphics[width=0.26\textwidth]{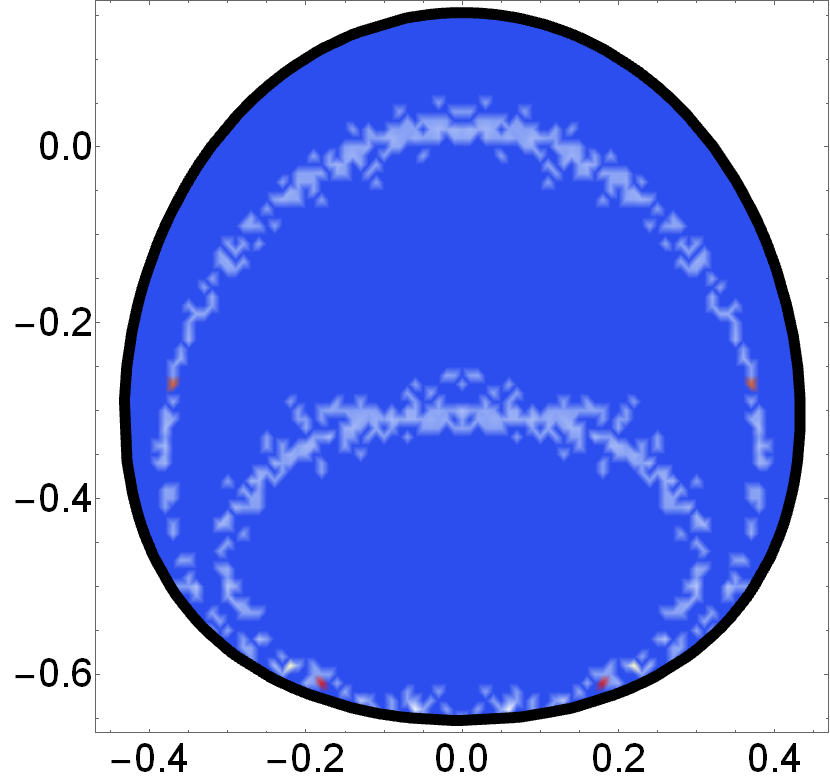}&\includegraphics[width=0.26\textwidth]{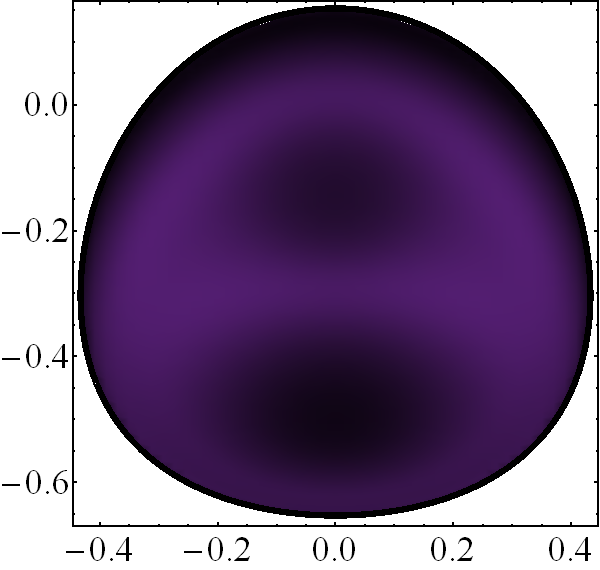} & $-1.8$\\
\scalebox{2}{$\tilde{j}_z$} &\includegraphics[width=0.26\textwidth]{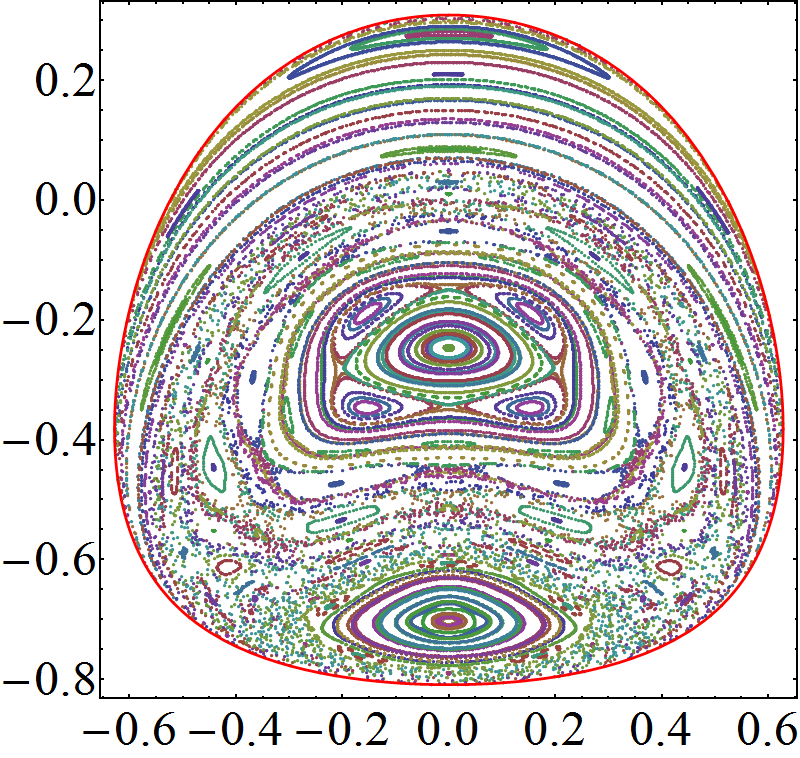}&\includegraphics[width=0.26\textwidth]{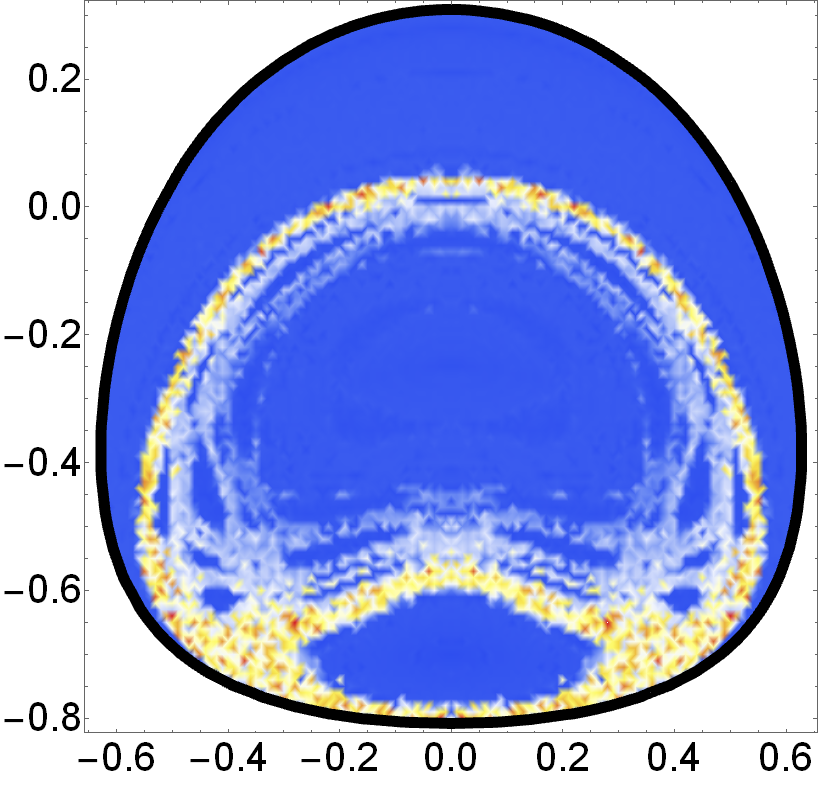}&\includegraphics[width=0.26\textwidth]{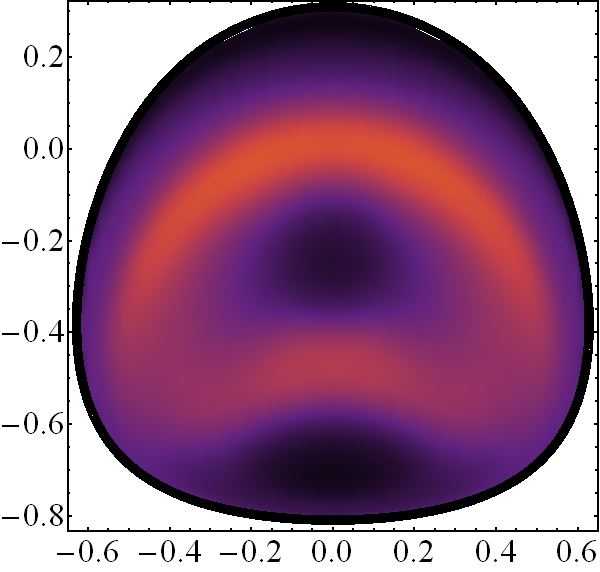}&$-1.5$ \\
& \includegraphics[width=0.26\textwidth]{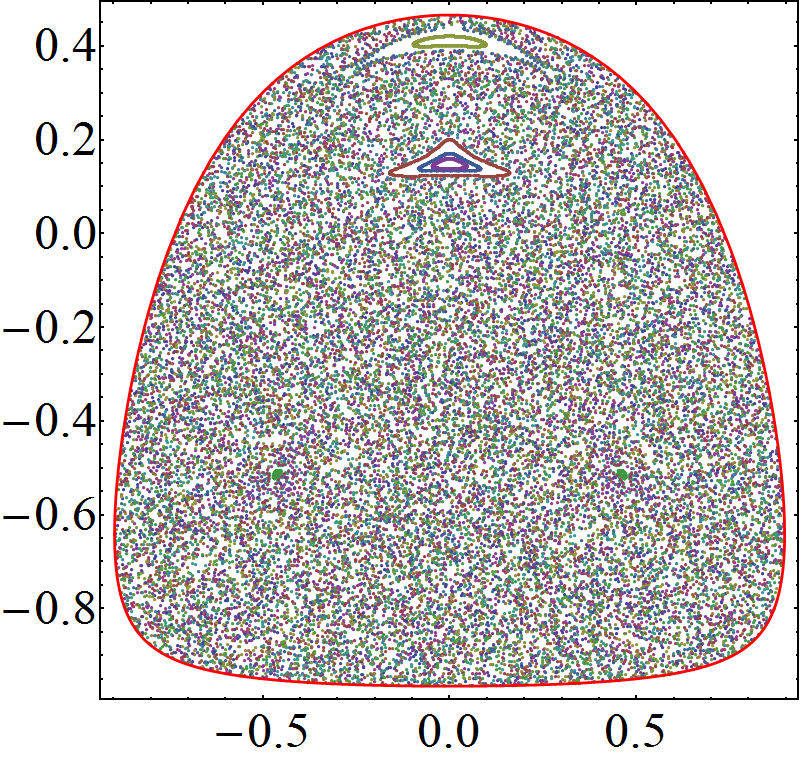}&\includegraphics[width=0.26\textwidth]{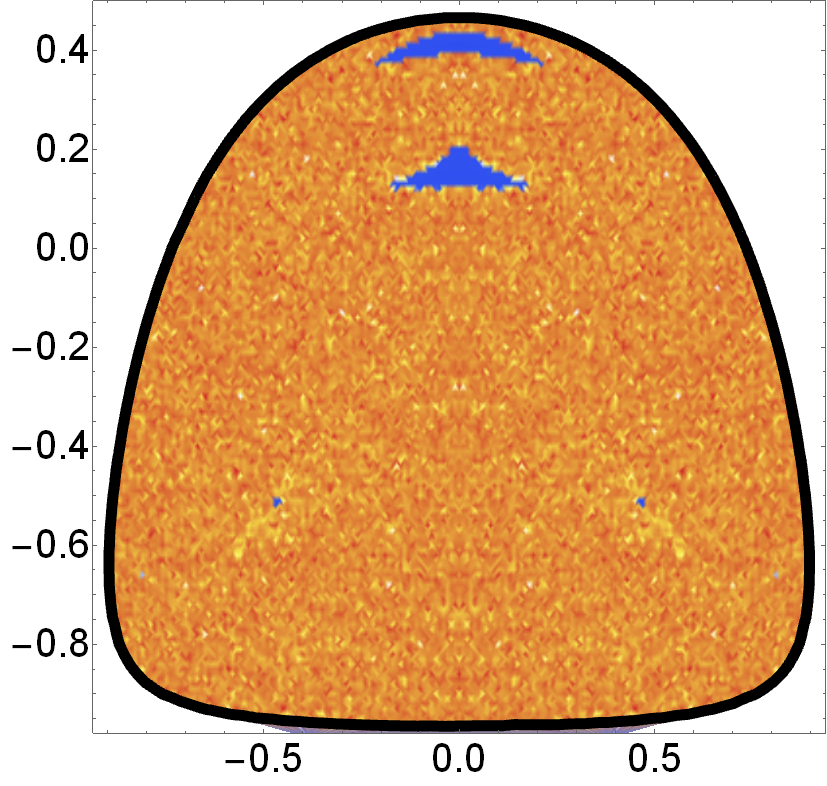}&\includegraphics[width=0.26\textwidth]{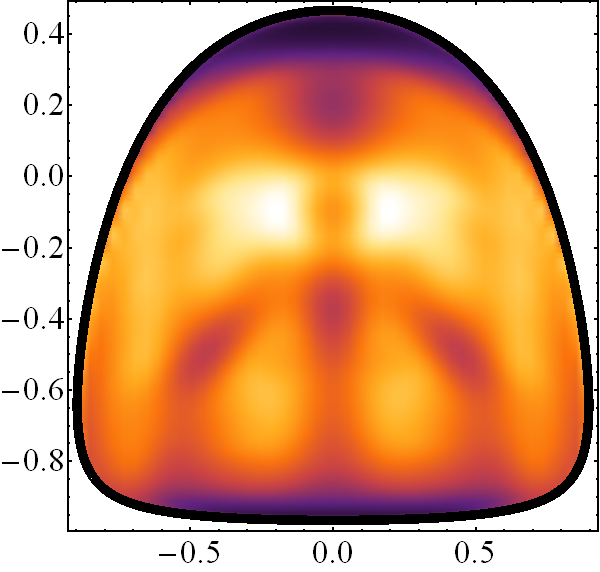}& $-1.1$\\
 & &\scalebox{2}{$\phi$} & & 
\end{tabular}
\caption{Poincar\'e sections (left column) and  Lyapunov exponents (central column) for a wide sample of initial conditions over the   Poincar\'e surface $p=0$, in the classical version of the Dicke model, for three different energies. The Poincar\'e surface is projected in the plane $\tilde{j_z}$ vs. $\phi$, with $\tilde{j_z}\equiv j_z/J$.   The  right column depicts  the Participation Ratio ($P_R$)  of quantum coherent states in the Dicke  Hamiltonian eigenbasis for $J=80$; the parameters of the coherent states were taken in the same Poincar\'e surface as in the classical results.  The Hamiltonian parameters  are $\omega=\omega_o=\gamma=1$. }
\label{fig:poin}
\end{figure*}

To investigate the presence of chaos in the classical version of the Dicke model, we solve numerically the Hamilton equations 
$$
\frac{d\phi}{dt}=\frac{\partial H_{D}^{cl}}{\partial j_z}, \ \ \ \ \ \frac{dj_z}{dt}=-\frac{\partial H_{D}^{cl}}{\partial \phi}, \ \ \ \ \
\frac{dq}{dt}=\frac{\partial H_{D}^{cl}}{\partial p},\ \ \ \ \ \frac{d p}{dt}=-\frac{\partial H_{D}^{cl}}{\partial q},
$$
and calculate from the numerical results Poincar\'e sections and maximal Lyapunov exponents. 
 
We consider a Poincar\'e surface in the phase space which is traversed by all  trajectories,  such a surface is $p=0$. Together with the energy conservation, the previous condition defines a two dimensional surface embedded in the 3D-space $q$-$j_z$-$\phi$. This surface is formed by two halves $q_{\pm}=q_{\pm}(E,j_z,\phi)$ coming from the two roots of the quadratic equation for $q$ imposed by energy conservation $H_D^{cl}(q,p=0,j_z,\phi)=E$.  To  visualize better the numerical results,  we select the upper  half, $q_+$, of the previous surface  and project it over the plane $j_z$-$\phi$.  

We choose an initial condition in the previous Poincar\'e surface and let it evolve. From the numerical trajectory we identify the points where the Poincar\'e surface is traversed at later times, obtaining a so-called Poincar\'e section. By repeating the previous procedure for a wide sample of initial conditions in the Poincar\'e surface, we are able to identify qualitatively the presence of chaos at a given energy by looking at the resulting Poincar\'e sections. When the points of a Poincar\'e section are organized in a regular pattern, the presence of a second (approximate) integral  of motion is revealed. In this case we say that the dynamics is regular. In contrast, if the points of a Poincar\'e section scatter in  an apparent random pattern, the non-integrability of the model is manifest and we have chaotic dynamics \cite{BasPRA142,BasPS15}.
 
To study quantitatively the presence of chaos, we calculate, also, the maximal Lyapunov exponent of a wide sample of initial conditions in  the same Poincar\'e surface. For this, we let evolve  many initial conditions located in a small circular neighbourhood around a central one. By measuring the separation of these trajectories   and averaging, we are able to estimate  the maximal Lyapunov exponent of the central initial condition (for details of the method see Ref. \cite{Cha16}). If the Lyapunov exponent is larger than a  numerical cut-off (typically 0.004), the separation is exponential and we have chaotic dynamics.  On the contrary, when  the Lyapunov exponent is very close to zero, we have a separation slower than exponential and thus regular dynamics. 

In  the left and central columns of Figure \ref{fig:poin}, we show results of the previous procedures  for different energies in the classical Dicke model with  $\omega=\omega_0=1$ and a coupling $\gamma=2\gamma_{cr}=1$ (the minimal energy for this case being $E_{GS}=-2.125 J$ and the critical energy of the ESQPT $E_{ESQPT}=-J$).  We can observe that the qualitative identification of regular and chaotic dynamics provided by the Poincar\'e sections (left column) is corroborated by the values of the Lyapunov exponents shown in central column. For low excitation energy the dynamics is predominantly regular, and as the energy increases chaotic dynamics is established in some regions of the phase space. For the highest energy shown, almost all the phase space is covered by chaotic dynamics.

For the quantum version of the Dicke model,   the unbounded number of bosons makes the Hilbert space infinite.  In order to diagonalize its Hamiltonian, a truncation in the  number of bosonic excitations is introduced. The cut-off has to be large enough to guarantee convergence of the low energy results we are interested in. We use the   basis described in \cite{Bastarrachea,Chinos}   to diagonalize  the Hamiltonian. This basis is particularly efficient to obtain, in the superradiant phase,  rapid convergence of a large portion of the low-energy spectrum  as a function of the cut-off. The values of $J$ computationally affordable are of order $J\sim 10^2$. We use  $J=80$ and $J=120$, and  consider the same case as in the classical results ($\omega=\omega_0=\gamma=1$).

In the search for a quantum signature that reveals the presence of chaotic dynamics in the classical limit, we consider initial coherent states. The quantum  dynamics of such initial states must be influenced by the kind of dynamics of the classical limit for large enough $J$.  We know that for any initial state $|\Psi_o\rangle$ the solution of the time-dependent Schr\"odinger equations is given by
$$
|\Psi, t\rangle= e^{-i \hat{H_D}t}|\Psi_o\rangle= \sum_{k} c_k e^{-i E_k t}|E_k\rangle,
$$ 
where $|E_k\rangle$ are eigenvectors of the Hamiltonian with energy $E_k$, and $c_k=\langle E_k|\Psi_o\rangle$. From this expression, it is clear that the decomposition of the initial state in the Hamiltonian eigenbasis codifies the temporal evolution of the initial state. A simple quantity giving information about the decomposition of the initial state in the Hamiltonian eigenbasis is the so called participation ratio $P_R$ defined as
$$
P_R= \frac{1}{\sum_k |c_k|^4}.
$$
This quantity estimates the number of Hamiltonian eigenstates participating in the unitary evolution of the initial state. Observe that if the initial state is stationary (eigenvector of the Hamiltonian), the $P_R$ is equal to one (provided  the energy levels have no degeneracies), while in the opposite limit if the initial state is uniformly  distributed along  $L$  Hamiltonian eigenstates,  $P_R=L$. Clearly, the evolution of the initial state depends also on the eigenenergies, but the participation ratio is simple to calculate and  provides information about the number of energies (and thus frequencies) contributing to the evolution of a given initial state.

In Refs. \cite{Bas16, BastaPS17, Haake, Zycz}, the participation ratio and  similar  quantities  were proposed   to detect chaos in initial coherent states of the kicked top and Dicke model. There,  numerical evidence was provided  of a direct relation between    the  $P_R$ (and quantities related) and chaos:  for coherent states with parameters defined in chaotic regions of the corresponding classical Hamiltonian the $P_R$ is large, whereas for coherent states in regular regions the $P_R$ is smaller. In the right column of Figure \ref{fig:poin}, we  show this evidence for the Dicke model. We plot the $P_R$ for many coherent states with parameters in the same Poincar\'e surface used to calculate the Poincar\'e sections and Lyapunov exponents of the classical version.
The figure makes evident the direct relation  between chaos in the classical version and the quantum $P_R$ for  coherent states. For classical regions with large Lyapunov exponent, the quantum coherent states show an increase of their $P_R$, and the opposite, coherent states in  classical regular regions have a low $P_R$.  The correspondence is not perfect, but this  is  expected because the quantum results are limited by the finite spreading of the coherent states in the phase space, which prevents to resolve structures with an area smaller than $2\pi/J$. For the same reason, the fine structures observed in the classical results could become detectable in the quantum results for large enough $J$.

\section{Quantum dynamics in  regular and chaotic  regimes}

The observed correspondence between classical chaos and the $P_R$ of coherent states in the Hamiltonian eigenbasis, can be further investigated by looking at the unitary evolution. The quantum dynamics must be influenced by the number of Hamiltonian eigenstates participating in the initial state, and also by the energies and distribution  of the participating Hamiltonian eigenstates. In this  section we focus on this issue, and identify the different properties of the distributions of the Hamiltonian eigenstates participating in initial coherent states located in regular and chaotic  regions of the corresponding classical phase space. In the case of regular dynamics, this study allows us to derive an analytical expression for the survival probability (SP). The survival probability, defined as 
\begin{equation}
\textrm{SP}(t)=\left| \langle \Psi(0) |\Psi(t) \rangle \right|^2 = 
\left|\displaystyle\sum\limits_{k} |c_k|^2 e^{-i E_k t}\right|^2 
= \displaystyle\sum\limits_{k \neq k'} |c_k|^2 |c_{k'}|^2 e^{-i(E_k-E_{k'})t}+ \frac{1}{P_R}, 
\label{SPdef}
\end{equation}
 is directly related to the $P_R$.  At long times, if the system does not have too many degeneracies, the sum in the right hand side of Equation~(\ref{SPdef}) averages to zero and $\textrm{SP}(t)$ just fluctuates around $1/P_R$. The dispersion of the temporal fluctuations of $\textrm{SP}(t)$ is also  $\sim 1/P_R$ \cite{TorresKollmar2015}.

In the following,  it is shown  that the small  $P_R$ obtained in the regular regions, can be understood because the   components of the energy eigenbasis participating in the coherent states have a recognizable pattern of one or several Gaussian distributions. On the other hand, the observed increase of the $P_R$ in  coherent states  within chaotic regions of the classical phase space  is  related with the absence of a simple structure in the Hamiltonian eigenbasis components of the coherent states.

\subsection{Regular region close to the onset of chaos}
\begin{figure*}
 \includegraphics[width=0.33\textwidth]{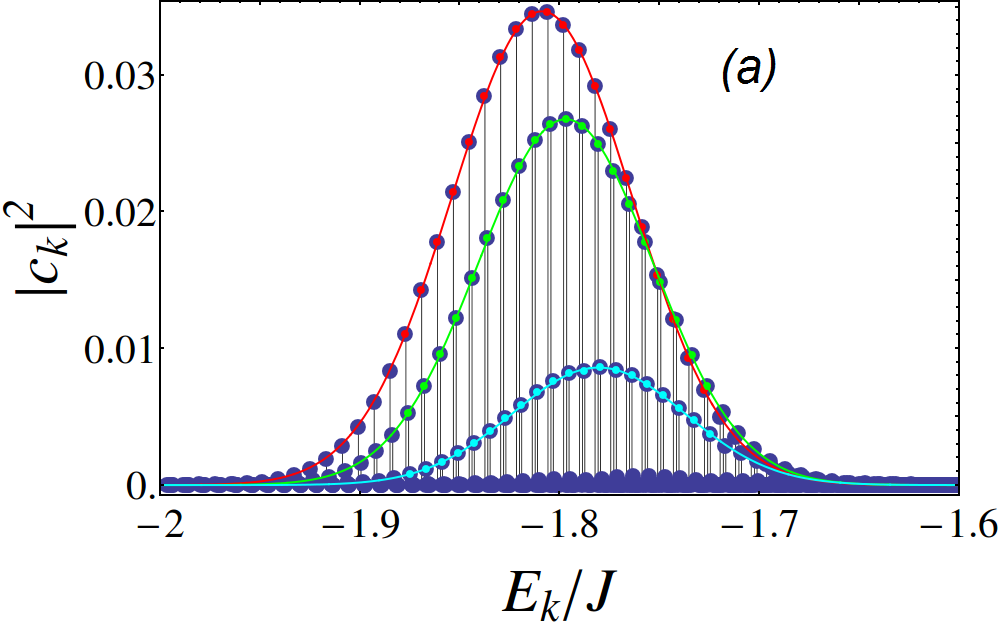} \includegraphics[width=0.23\textwidth]{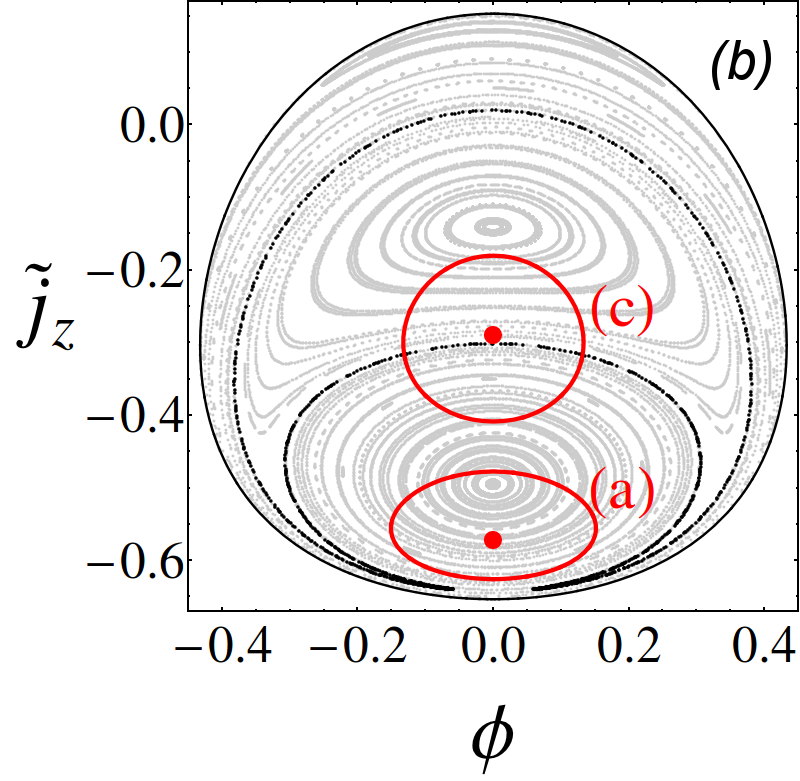}\ \ \ \includegraphics[width=0.34\textwidth]{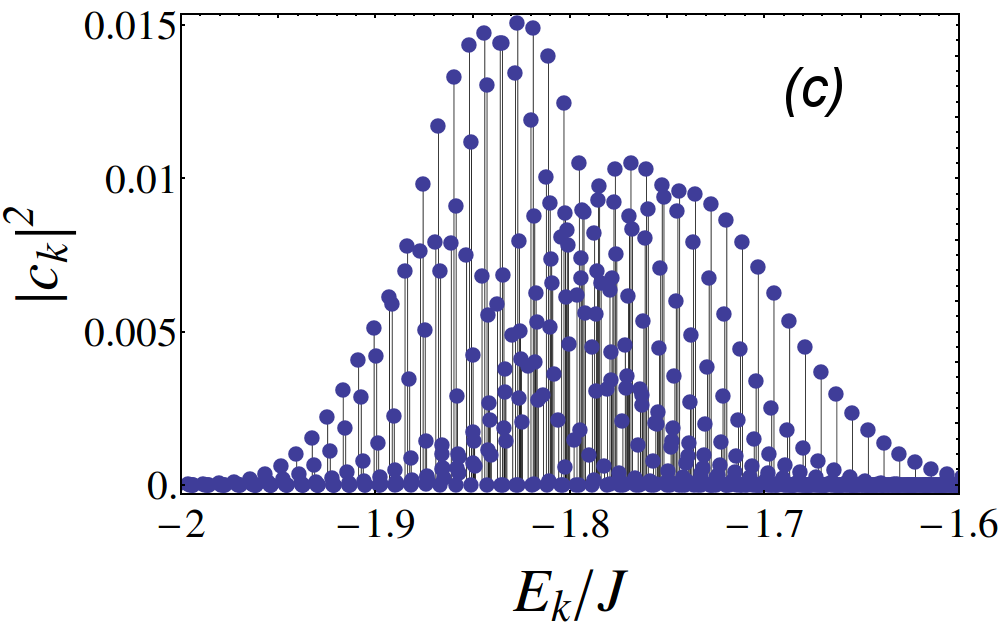}\\
 \begin{tabular}{cc}
 \includegraphics[width=0.45\textwidth]{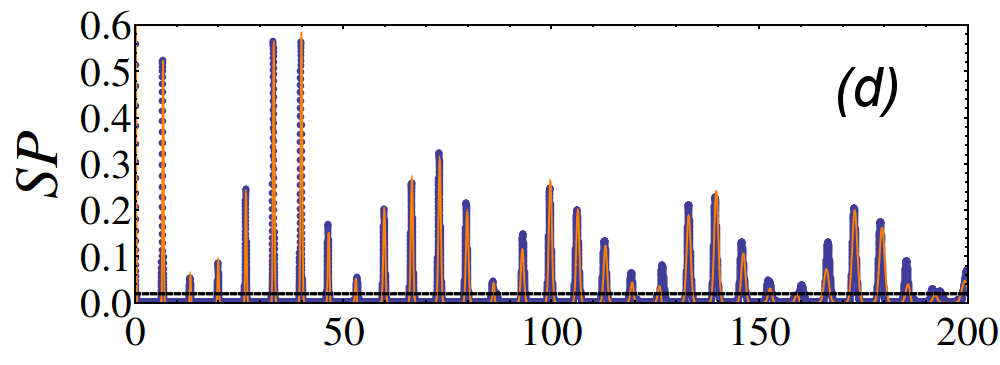}&\includegraphics[width=0.45\textwidth]{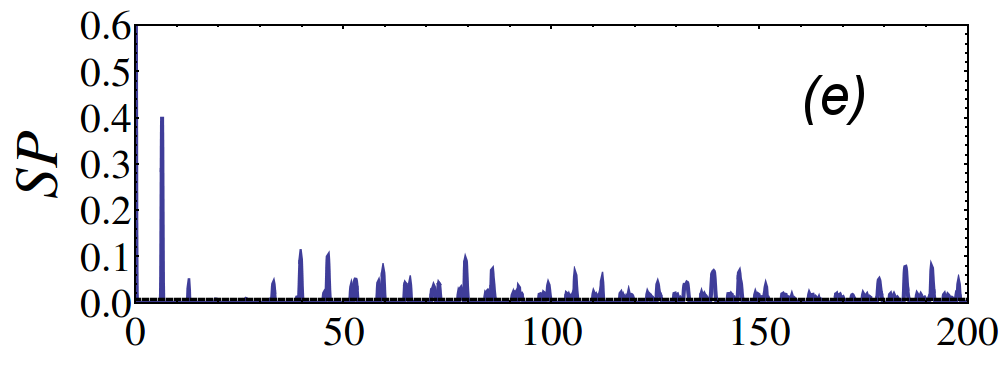}\\
{\Large\textit{t}} &  {\Large\textit{t}}
 \end{tabular}
\caption{Absolute squared components, $|c_k|^2$, of two coherent states [(a) and (c)] in the Hamiltonian eigenbasis with mean energy $E=-1.8 J$ and $J=120$. Continuous lines  in panel (a) are Gaussian fits to the the components of three sequences of energy states participating in the coherent state. The localization and spreading of the coherent states in the classical phase space are depicted in panel (b).  Panels (d) and (e) show the survival probability of the initial coherent states of panels (a) and (c) respectively. Dark blue lines are numerical results and light orange line in panel (d) is the analytical approximation in Equation (\ref{Eq:SPwithInt}), (\ref{Eq:iscon}) and (\ref{Eq:incon}). Horizontal lines in panels (d) and (e) show the respective values of $1/P_R$. In panel (e), since $P_R$ is large,  this line is hardly distinguishable from the horizontal axis. }
\label{fig:locat}
\end{figure*}

In this section we show results for the survival probability of selected initial coherent states at energy $E=-1.8 J$, where regular dynamics dominates the classical phase space. We select two  coherent states, the localizations of their   parameters,  in  the same Poincar\'e surface of first row in Figure \ref{fig:poin}, are shown  in Figure \ref{fig:locat}(b)  with dots. The curves surrounding  the dots  depict the spreading of the initial coherent state in the phase space. They were obtained by calculating the overlap $|\langle z\alpha|z_o \alpha_o\rangle|^2$ and identifying the points where this overlap decays to $e^{-1}$.  In order to correlate the quantum dynamics of the coherent states with  the corresponding  underlying classical dynamics, the  Poincar\'e sections are shown beneath  in gray scale.  The decomposition of the two chosen  coherent states in terms of the Hamiltonian eigenbasis is shown in  Figure \ref{fig:locat}(a) and \ref{fig:locat}(c). For the regular state of  Figure \ref{fig:locat}(a),  the components are organized in several sequences, distributed each according to Gaussian distributions $ |c_k^{(i)}|^2\approx g_{k}^{(i)}=A_i \exp\left[-\frac{(E_{k}^{(i)}-\bar{E}_i)^2}{2 \sigma_i^2}\right]$,  whose amplitudes ($A_i$), widths ($\sigma_i$) and means ($\bar{E}_i$) can be estimated by the fittings shown in the same figure by continuous lines. The index $i$  runs from $1$ to  the number of identifiable sequences $M$ ($M=3$ in the case shown). In contrast, for the state in Figure \ref{fig:locat}(c), which is located in the separatrix between two sets of regular trajectories,  where according to Figure \ref{fig:poin}   the Lyapunov exponent acquires a value different to zero,  the energy components show  a more intricate structure and the $P_R$ is larger.

For regular coherent states, as mentioned before,  the observed structure of their energy components allows to derive an analytical expression for the SP.  By identifying  the components and eigenenergies of each sequence by $ c_k^{(i)}$ and $E_{k}^{(i)}$, the SP can be written as
\begin{eqnarray}
&&\textrm{SP}(t)=\left|\sum_{ik} |c_k^{(i)}|^2 e^{-i E_k^{(i)}t}\right|^2 
= \sum_{i}\sum_{kk'}2|c_k^{(i)}|^2|c_{k'}^{(i)}|^2\cos\left[\left(E_{k'}^{(i)}-E_{k}^{(i)}\right)t\right] +\sum_{i<j}\sum_{kk'} 2 |c_k^{(i)}|^2|c_{k'}^{(j)}|^2\cos\left[\left(E_{k'}^{(j)}-E_{k}^{(i)}\right)t\right]\nonumber\\
&&=\sum_{i} \textrm{SP}^{(i)}(t)+\sum_{i<j} \textrm{SP}_I^{(ij)}(t).
\label{Eq:SPwithInt}
\end{eqnarray}
In reference \cite{Lerma17},  it is shown that  for each sequence
\begin{equation}
\textrm{SP}^{(i)}(t) = \frac{A_i^2 \sigma_i\sqrt{\pi}}{\omega_1^{(i)}} \Theta_3(x,y), \label{Eq:iscon} 
\end{equation}
where  $x=\omega_1^{(i)} t/2$ and $y=\exp\left[-\frac{1}{4}\left(\frac{\omega_1^{(i)}}{\sigma_i}\right)^2\right] \exp\left[-\left(\frac{t}{t_D^{(i)}}\right)^2\right]$, with $\Theta_3(x,y)= 1+2\displaystyle\sum_{p=1} y^{p^2}\cos(2px)$,  the Jacobi Theta function.  The parameters entering in the previous definitions  are: a)   the frequency $\omega_1^{(i)}=E_{kmax+1}^{(i)}-E_{kmax}^{(i)}$ is the  difference of the closest energies of the $i$-th sequence to the mean energy $\bar{E}_i$ ($E_{kmax}^{(i)}\leq \bar{E}_i \leq E_{kmax+1}^{(i)}$); b)  the anharmonicity  $e_2^{(i)}= (E_{kmax+1}^{(i)}+E_{kmax-1}^{(i)})/2-E_{kmax}^{(i)}$;  and  c) the decay time of each isolated sequence  $ t_D^{(i)}\equiv \frac{\omega_1^{(i)}}{|e_2^{(i)}|\sigma_i}$.
 
In the same reference, it is shown that   the interference term between $i$-th and $j$-th sequences is  
\begin{equation}
\textrm{SP}_I^{(ij)}(t)=  \frac{2 A_i A_j\sqrt{2\pi}\sigma_i\sigma_j}{\omega_{ij}\sqrt{\sigma_i^2+\sigma_j^2}} 
\displaystyle\sum_{p\in \mathbb{Z}} \exp\left[{-\frac{(p \omega_{ij}+\delta E_{ij}+\bar{E}_i-\bar{E}_j)^2}{2(\sigma_i^2+\sigma_j^2)}}\right]\exp\left[{-\frac{(\sigma_{ij} p t)^2}{2}}\right]\cos[(\delta E_{ij}+p\omega_{ij})t],
\label{Eq:incon} 
\end{equation}
where
$
\sigma_{ij}=\frac{2 |e_2^{(i)}| \sigma_i\sigma_j}{\omega_{ij} \sqrt{\sigma_i^2+\sigma_j^2}},
$
$\delta E_{ij} =\langle E_{k}^{(i)} - E_{k}^{(j)}\rangle$  is the mean energy difference between the energies of sequence $i$ and $j$, and  
$\omega_{ij}=E_{k_I+1}^{(i)}-E_{k_I}^{(i)}$ is the energy difference between the eigenvalues of the $i$-th sequence that are closest to the value $E^{(I)}_{ij}$ that maximizes the product of the Gaussians $g_{k}^{(i)} g_{k}^{(j)}$. This latter value is given by   
$
E^{(I)}_{ij}=\frac{\bar{E}_i \sigma_j^2 + \bar{E}_j \sigma_i^2}{\sigma_i^2 + \sigma_j^2}
$
and satisfies $E_{k_I}^{(i)}\leq E^{(I)}_{ij}\leq E_{k_I+1}^{(i)}$. 

In Figure \ref{fig:locat}(d)  we compare the previous approximation with the numerical results for the SP of the regular  coherent state of Figure \ref{fig:locat}(a). The agreement is remarkable and the approximation describes very well the numerics from $t=0$ up to the time 
 where the survival probability begins to fluctuate around  $1/P_R$.  In Figure \ref{fig:locat}(e), we show the  numerically   calculated SP for the coherent state of Figure \ref{fig:locat}(c). As mentioned, this state is located in a region with a Lyapunov exponent different to zero,  and this is reflected by a rapid decay  of the SP to the regime of fluctuations around $1/P_R$, with a pattern completely different to the obtained in the  previous regular case.    
 
\begin{figure*}
 \includegraphics[width=0.34\textwidth]{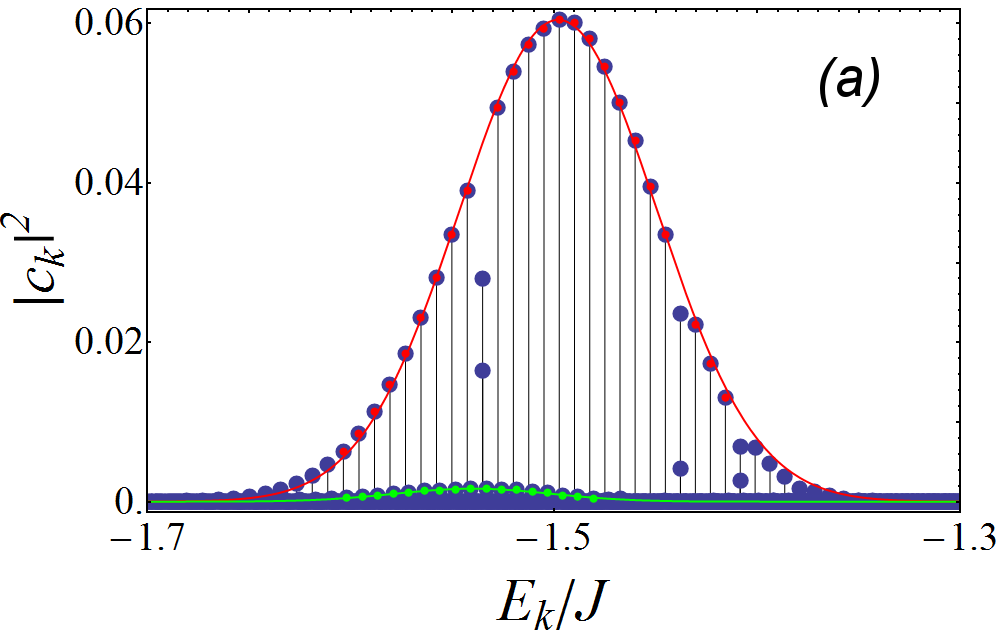} \ \ \ \includegraphics[width=0.23\textwidth]{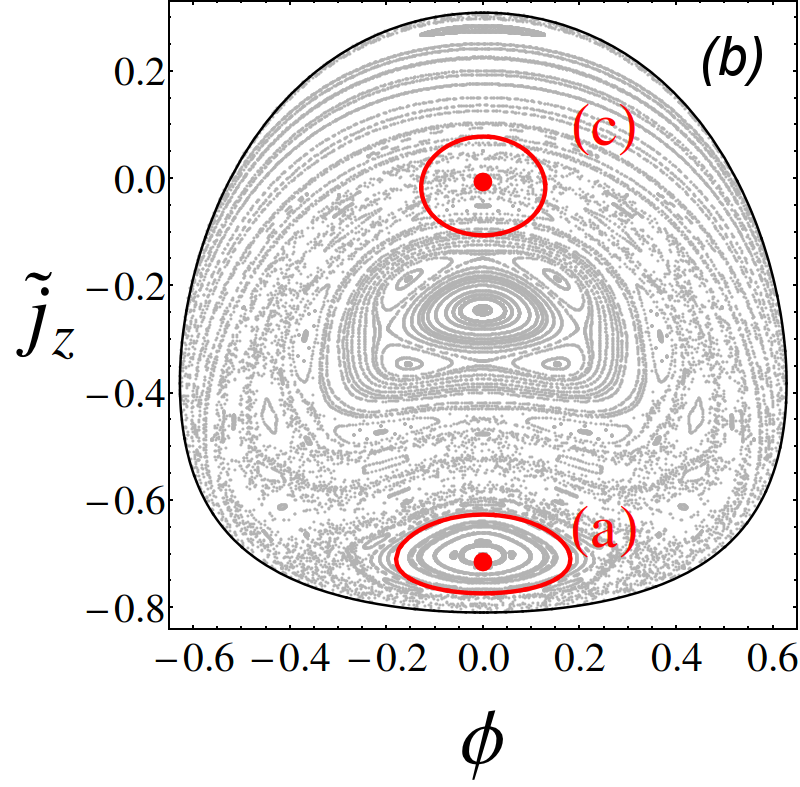} \includegraphics[width=0.34\textwidth]{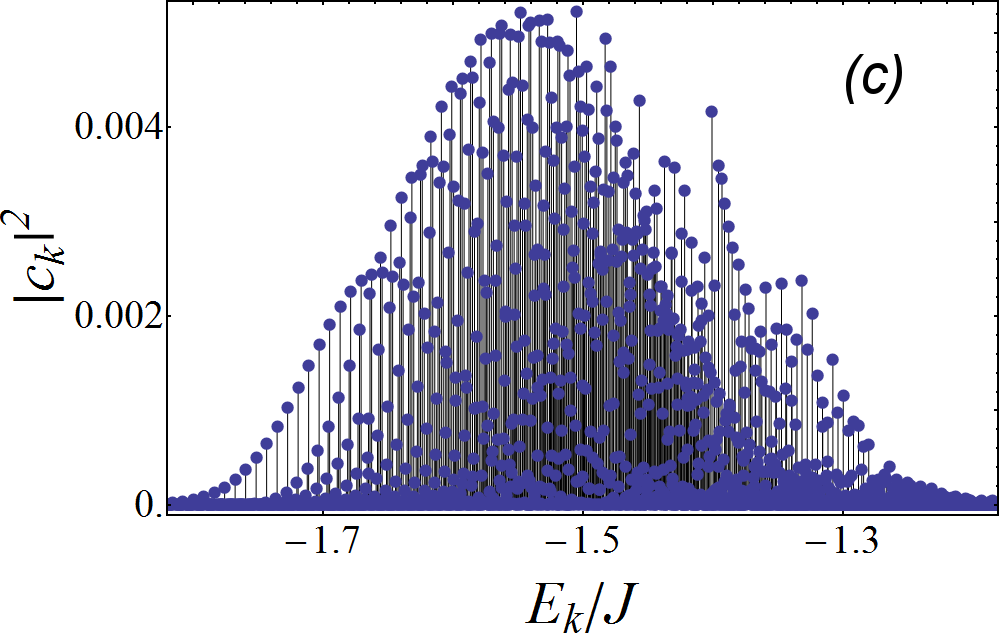} \\
 \begin{tabular}{cc}
 \includegraphics[width=0.45\textwidth]{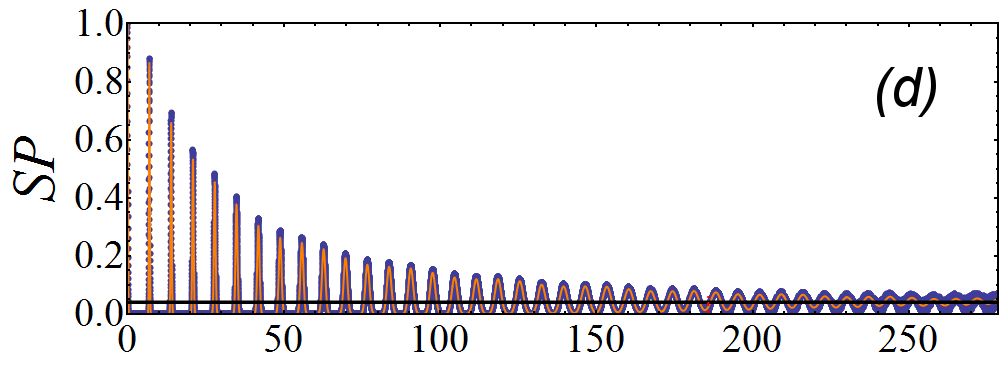}&\includegraphics[width=0.45\textwidth]{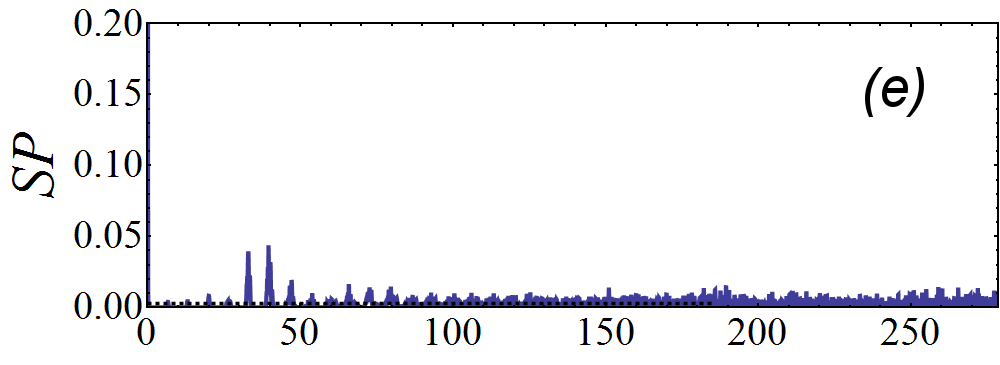}\\
{\Large\textit{t}} & {\Large\textit{t}}
\end{tabular}
\caption{Same as Figure \ref{fig:locat} but for a higher mean energy, $E=-1.5 J$, where regular and chaotic regions coexist. Coherent states (a) and (c) are located, respectively,  in a regular and a chaotic classical region. In order to visualize the weak revivals of the SP of coherent state (c), a smaller vertical scale was used in panel (e). }
\label{fig:em1p5}
\end{figure*}

\subsection{Mixed and chaotic regions}

In this section we further study the correspondence between the value of  the $P_R$ and the temporal evolution of the SP in  coherent states defined in regular and chaotic regions at larger energies.  In Figure \ref{fig:em1p5}(a) and \ref{fig:em1p5}(c)  the energy components of two coherent states at energy $E=-1.5 J$ are displayed. Their  localizations in  phase space are shown  in Figure \ref{fig:em1p5}(b). The  coherent states \ref{fig:em1p5}(a) and \ref{fig:em1p5}(c)  are  located in  regular and chaotic  regions respectively  of the underlying classical phase space. As in the previous lower energy case, the structure of the components in the regular case shows  sequences of components distributed according to Gaussians, whereas in the chaotic case no identifiable structure  can be recognized. For the regular case the analytical expression in Equation (\ref{Eq:SPwithInt}) is applicable. The comparison between the numerically calculated SP and the analytical approximation is shown in Figure \ref{fig:em1p5}(d), where, again, a  very good accord is obtained. In Figure \ref{fig:em1p5}(e) the SP of the chaotic initial coherent state confirms what was observed  previously: the SP decays rapidly to the  regime of fluctuations, following a  different pattern respect  to the regular cases.  

\begin{figure*}
\includegraphics[width=0.34\textwidth]{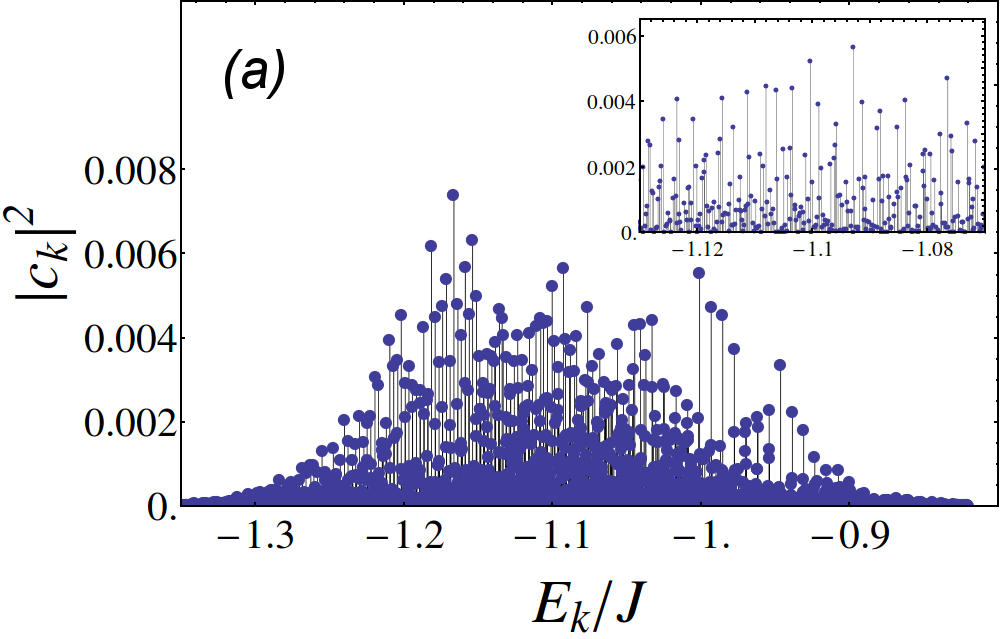} \ \ \ \includegraphics[width=0.23\textwidth]{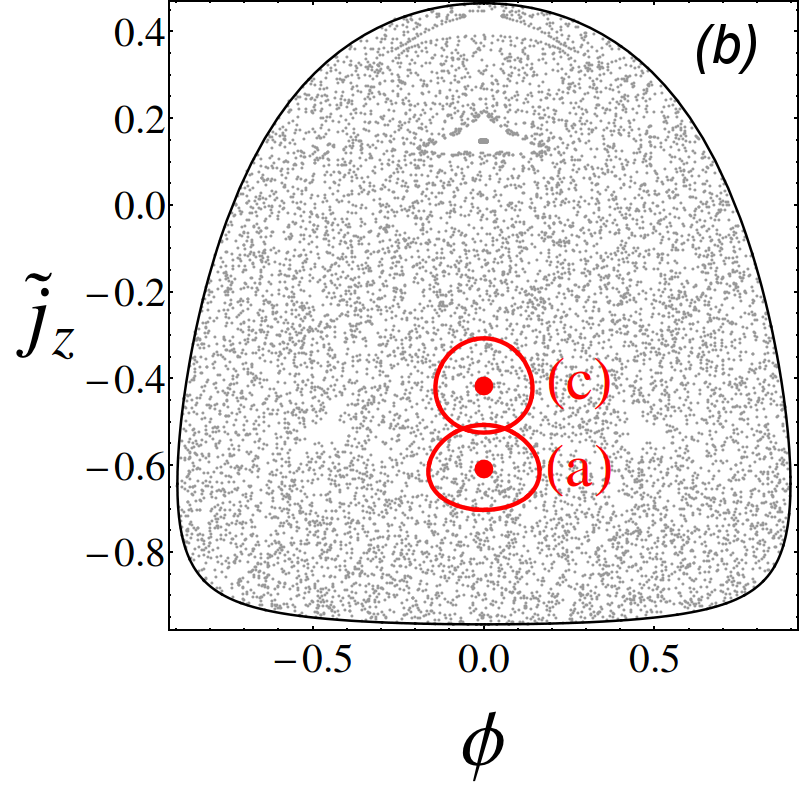} \includegraphics[width=0.34\textwidth]{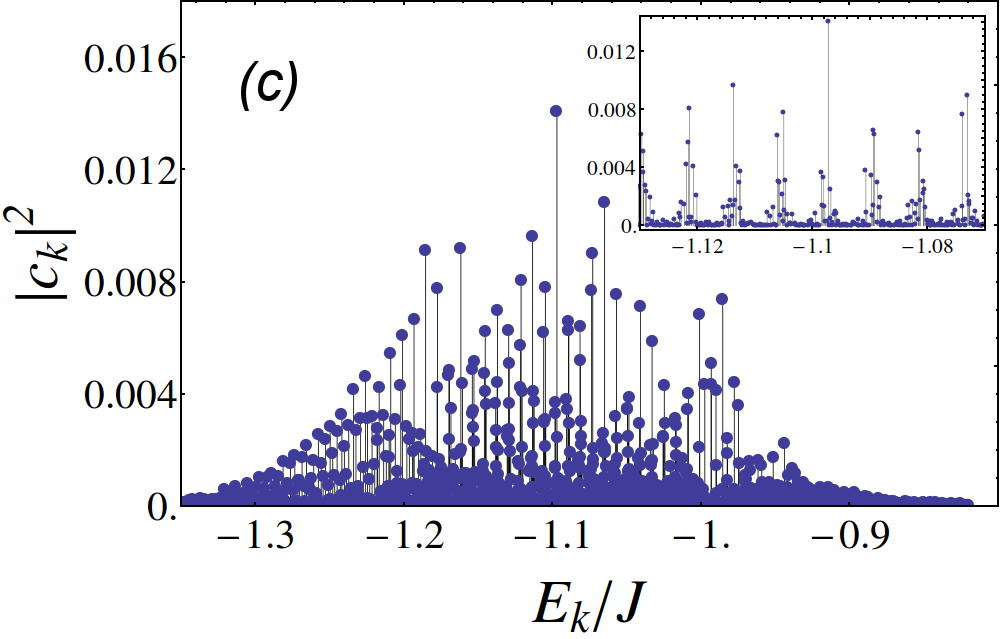} \\
\begin{tabular}{cc}
 \includegraphics[width=0.45\textwidth]{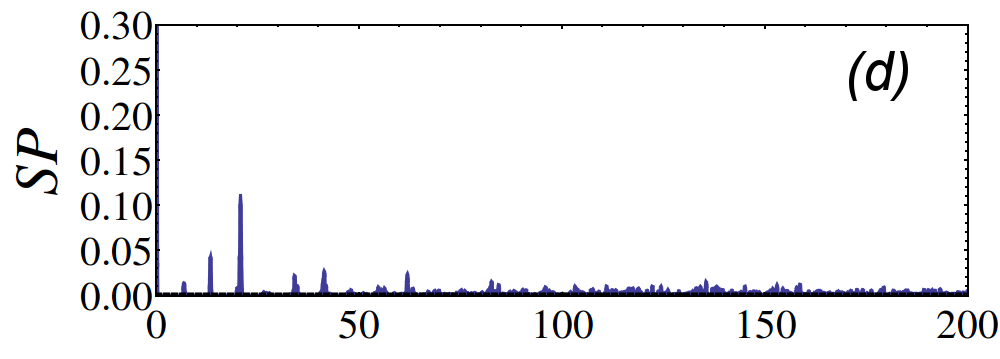}&\includegraphics[width=0.45\textwidth]{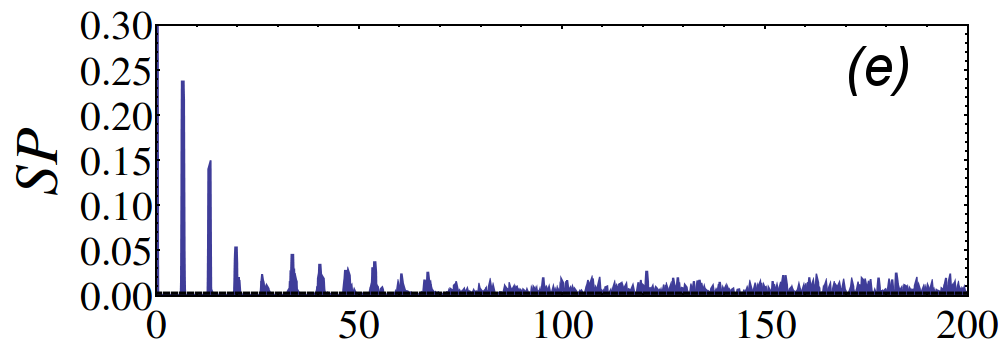}\\
{\Large\textit{t}} & {\Large\textit{t}}
\end{tabular}
\caption{Same as Figure \ref{fig:locat}, but for mean energy $E=-1.1 J$, where the classical phase space is covered almost completely by chaotic trajectories. Both coherent states (a) and (c) are located in the chaotic region. The insets in panels (a) and (c) are a closer view of the energy components in a smaller energy scale.}
\label{fig:em1p1}
\end{figure*}

In Figure \ref{fig:em1p1}(a) and \ref{fig:em1p1}(c) the components   of two coherent states at energy $E=-1.1 J$ are shown. Their localizations in the phase space are depicted in  Figure \ref{fig:em1p1}(b). At this energy, the phase space is covered almost entirely by chaotic trajectories. This is reflected by the components of the coherent states, which do not show a structure of Gaussian distributed energy sequences. However, a  view of the components in a smaller energy scale [see insets in Figure \ref{fig:em1p1}(a) and \ref{fig:em1p1}(c)], reveals that in the case of coherent state \ref{fig:em1p1}(c) the components have regularities absent in the case of coherent state \ref{fig:em1p1}(a).  The  SP [shown in Figure \ref{fig:em1p1}(d) and \ref{fig:em1p1}(e)] decays rapidly in both cases, but due to its partial regular structure,   the SP of coherent state \ref{fig:em1p1}(c) shows a few relatively large partial revivals, which are not present in the case of coherent state \ref{fig:em1p1}(a). This difference is related to the scarring phenomenon, already identified in \cite{Heller}, which occurs when the initial coherent state is located close to an unstable classical periodic orbit, which we conjecture is the case of coherent state of Figure \ref{fig:em1p1}(c).

\section{Conclusions}

Numerical evidence was provided, showing that the number of participating Hamiltonian eigenstates in coherent states  is a quantity sensitive to the presence of chaos in the semi-classical limit of the Dicke model. 
The dynamics of  coherent quantum states in the semi-classical limit of the Dicke model was studied 
employing the Survival Probability. An analytical expression for this quantity in terms of the Jacobi theta function was provided  for coherent states in regular regions of the classical phase space. 
The analytical expression describes remarkably well the numerical results from the initial time until the time when the SP begins to fluctuate around its asymptotic value ($1/P_R$). The analytical expression was obtained by noticing that    the energy levels participating in  the initial  coherent state  are organized in one or several sequences with components  described by  Gaussian distributions.

The way as classical chaos  influences the distribution of the initial coherent state in terms of the energy eigenstates was also discussed. For coherent states  in chaotic regions  a distribution with no simple structure was observed.  This is reflected in the temporal behavior of the SP  by weak revivals and a quick establishing of the asymptotic (equilibration) regime of fluctuations  around  $1/P_R$.  
 
The analysis presented here can be extended to study the dynamical evolution of other quantum observables, and although we use the Dicke  model to illustrate our approach, it is valid in general models with few degrees of freedom.     

\section{ACKNOWLEDGMENTS}
We are very grateful to Lea Santos for fruitful discussions. We acknowledge financial support from Mexican CONACyT project CB2015-01/255702,  DGAPA-UNAM project IN109417 and RedTC. MABM is a post-doctoral fellow of CONACyT. JChC and BLC acknowledge financial support from CONACyT and SNI-CONACyT scholarship programs. SLH acknowledges financial support from  CONACyT fellowship program for sabbatical leaves.

\bibliographystyle{aipnum-cp}%

\end{document}